\documentclass[conference]{IEEEtran}
\IEEEoverridecommandlockouts
\usepackage{cite}
\usepackage{amsmath,amssymb}
\usepackage{booktabs,tabularx,array,multirow}
\usepackage{graphicx}
\usepackage{xcolor}
\usepackage{tikz}
\usetikzlibrary{arrows.meta,positioning,calc,fit,shapes.geometric}
\usepackage{algorithm}
\usepackage{algpseudocode}
\usepackage{pifont}
\usepackage{microtype}
\usepackage{xspace}
\usepackage{balance}
\usepackage[hidelinks]{hyperref}

\usepackage{fancyhdr}
\fancypagestyle{paperstyle}{
    \fancyhf{}
    \fancyhead[L]{\footnotesize \textit{Preprint. To appear at ICCD 2026, 16-18 November 2026, Hong Kong, China.}}
    \fancyhead[R]{\footnotesize \thepage}
    
}

\DeclareRobustCommand{\circled}[1]{\tikz[baseline=(char.base)]{
    \node[shape=circle,draw,inner sep=0.2pt] (char) {\tiny #1};
  }}

\newcommand{\system}{\textsc{CONFERM}\xspace}

\newcolumntype{Y}{>{\raggedright\arraybackslash}X}
\newcolumntype{C}[1]{>{\centering\arraybackslash}p{#1}}
\newcolumntype{L}[1]{>{\raggedright\arraybackslash}p{#1}}

\newcommand{\jyin}[1]{\textcolor{black}{#1}} 
 
\newcommand{\cfang}[1]{\textcolor{black}{#1}}

\newcommand{\mv}[1]{\textcolor{black}{#1}}
\newcommand{\jyincamrdy}[1]{\textcolor{black}{#1}} 

\begin{document}

\title{CONFERM: Recurrence-Aware Temporal Mapping for Multi-Cycle Multi-Context CGRAs}

\author{
\IEEEauthorblockN{  Jun~Yin,
                    Jannes~Willemen,
                    Stef~Cuyckens,
                    Chao~Fang,
                    Marian~Verhelst}
\IEEEauthorblockA{\textit{ESAT-MICAS KU Leuven}}
\thanks{This project has been partly funded by the European Research Council (ERC) under grant agreement No. 101088865, the Flanders AI Research Program, and long-term structural Methusalem funding by the Flemish Government. \textit{Email: \{jun.yin, jannes.willemen, stef.cuyckens, chao.fang, marian.verhelst\}@kuleuven.be. Corresponding author: Chao Fang.}}}

\maketitle

\setlength{\intextsep}{6pt plus 2pt minus 1pt}
\setlength{\textfloatsep}{12pt plus 2pt minus 2pt}
\setlength{\floatsep}{12pt plus 2pt minus 2pt}
\setlength{\abovecaptionskip}{3pt plus 1pt minus 1pt}
\setlength{\belowcaptionskip}{2pt plus 1pt minus 1pt}
\setlength{\dbltextfloatsep}{10pt plus 2pt minus 2pt}
\raggedbottom

\pagestyle{paperstyle}
\thispagestyle{paperstyle}

\begin{abstract}
\cfang{Throughput in DSP and machine learning workloads is often limited by two temporal structures, i.e., loop-carried recurrences and long-latency, multi-cycle compute \mv{nodes}.}
On \jyincamrdy{spatio-temporal} coarse-grained reconfigurable arrays~(CGRAs), both bottlenecks can be addressed by overlapping iterations across \jyincamrdy{the multi-context modulo configurations}.
\mv{Yet, existing \cfang{CGRA} mappers schedule a \jyincamrdy{fixed} \cfang{dataflow graph (DFG)} that \jyincamrdy{treats recurrence-aware scheduling and operator-level pipelining separately}, limiting inter-iteration overlap and inflating routing pressure. 
To tackle this, we present CONFERM, a recurrence-aware temporal mapper that \jyincamrdy{uses} the dominant temporal constraint to guide the DFG representation and expose opportunities for loop-carried pipelining. 
CONFERM identifies \jyin{\jyincamrdy{and prioritizes bottleneck regions} during scheduling. 
The \mv{regular} loop-carried offsets across interleaved iterations allow the emitted control sequence to repeat at a shorter cadence than the \mv{original} initiation interval, thus delivering higher throughput with lower CGRA configuration overhead.}
Across ten benchmark kernels, CONFERM improves throughput by 2.18$\times$ over state-of-the-art mappers. Its \jyincamrdy{uniform} iteration offsets shorten the emitted initiation interval by 46\%. CONFERM's mapper pass also converges faster by 5.07$\times$ on average with the same heuristic mapper backend.}

\end{abstract}


\section{Introduction}

\begin{figure}[t]
  \centering
  \includegraphics[width=\columnwidth]{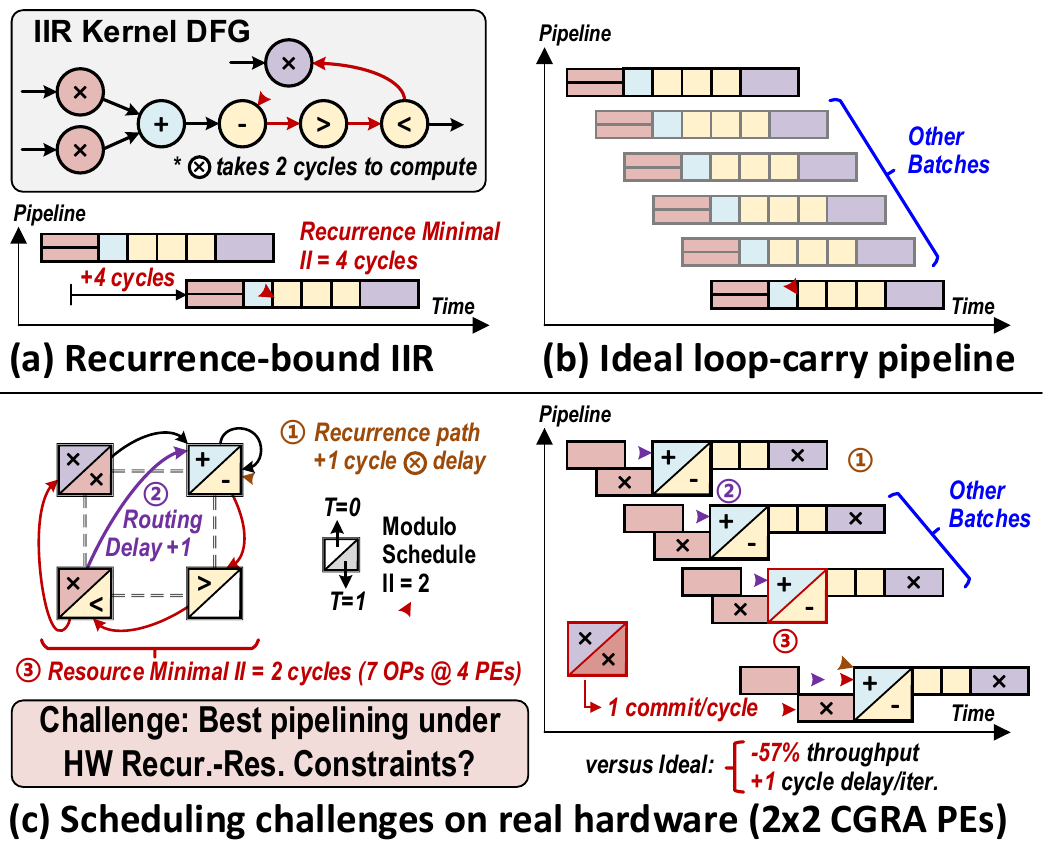}
  \caption{Overview of the background and challenges. \textbf{(a)} Recurrence-bound kernels (e.g. IIR~\cite{oppenheim1989discrete}) limit the achievable throughput by enforcing the loop-carried feedback dependency in general modulo schedulers~\cite{rau1994ims}; \textbf{(b)} The ideal throughput improvement can be achieved via loop-carried pipelining over independent data in the loop or over batches; \textbf{(c)} On real hardware, the limited  routing \textbf{\circled{2}} and execution \textbf{\circled{3}} resources delay operations and introduce pipeline bubbles, reducing throughput relative to the ideal schedule
  The real multi-context CGRA hardware constraints must therefore be considered when constructing a feasible interleaved schedule. 
  }
  \label{fig:overview}
\end{figure}

\cfang{Throughput in DSP and machine learning workloads~\cite{oppenheim1989discrete, gu2024mamba, dao2022flashattention} is often bounded not by the full loop body but by two recurring temporal structures. One is a loop-carried recurrence, where an operation in one iteration depends on a value produced by an earlier one, as in infinite impulse response (IIR) filters~\cite{oppenheim1989discrete}, whose kernel dataflow graph is shown in Fig.~\ref{fig:overview}(a).
The other is a long-latency, multi-cycle compute \mv{node}, where a single operator can take several cycles to produce a result, as in the nonlinear activation layers of modern machine learning models~\cite{zhao2025quark, du2025algorithm, qin2025picachu}. In both cases, \mv{the throughput of the looped workload} depends on whether the hardware executing it can keep multiple iterations in flight across time.}
\jyincamrdy{
Under modulo scheduling~\cite{rau1994ims}}, \cfang{the performance of a mapped loop is primarily measured by the initiation interval (II), the number of clock cycles between the launches of successive loop iterations. A lower II directly translates to higher throughput.}
\cfang{The two temporal structures above are \jyincamrdy{can prevent further reductions in II}.}
In the IIR kernel of Fig.~\ref{fig:overview}(a), the feedback path alone sets a four-cycle bound on II.
\jyincamrdy{Although ideal loop-carry pipelines like Fig.~\ref{fig:overview}(b) can be achieved by interleaving other batches, the achievable pattern is limited by available compute and routing resources.}

Coarse-grained reconfigurable arrays (CGRAs)~\cite{tan2020opencgra, qin2025picachu, yin2026stella, koul2024onyx, koul2023aha, prabhakar2017plasticine} are \jyincamrdy{compute architectures with highly flexible operator support} \cfang{that meet exactly this demand}.
\cfang{Unlike a \jyincamrdy{dedicated accelerator, CGRAs let} compilers map a loop directly onto a grid of \jyincamrdy{flexible} processing elements (PEs), \jyincamrdy{forming a tightly coupled data stream.}} \cfang{Yet, as Fig.~\ref{fig:overview}(c) shows, the seven operations of the loop body still set a separate two-cycle bound from resource demand on an array with only four PEs that can each start one operation per cycle.}
\jyincamrdy{To overcome a mapping failure, multi-context spatio-temporal CGRAs ~\cite{mei2003modulo, tan2020opencgra, tan2023vecpac} mitigate these constraints by allowing each PE to cycle through multiple configurations according to a modulo schedule (Fig.~\ref{fig:overview}(c)).}

\begin{figure*}[!t]
  \centering
  \includegraphics[width=.95\linewidth]{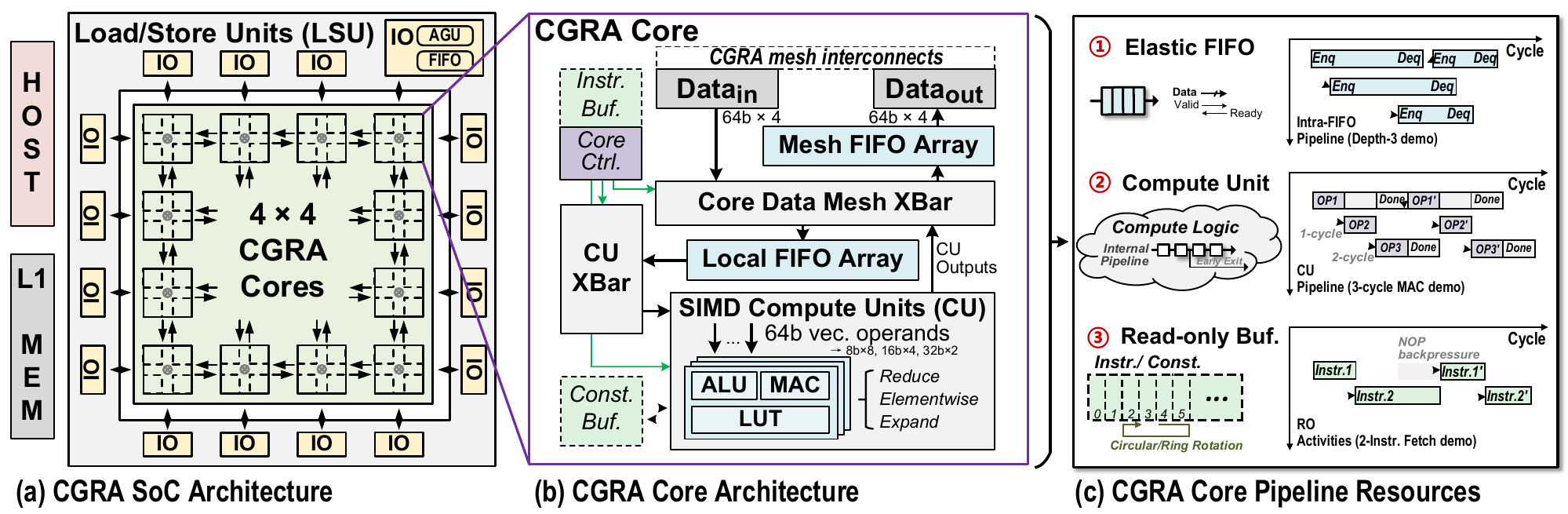}
  \caption{\jyincamrdy{The target spatio-temporal multi-context CGRA architecture}. \textbf{(a)} A $4\times4$ mesh of compute tiles, drawn as CGRA cores, with border load/store units that stream operands. \textbf{(b)} \jyincamrdy{Each core contains} SIMD CUs, local and mesh FIFO arrays, crossbars, and instruction/constant buffers. \jyincamrdy{The instruction buffers store the} modulo schedule locally. \textbf{(c)} The three pipeline resources that form the temporal depth of a core, namely elastic ready/valid FIFOs, internally pipelined CUs, and read-only buffers with a cyclic schedule. \system leverages these \jyincamrdy{pipeline resources to map recurrences} and critical feed-forward paths efficiently.}\label{fig:arch}
\end{figure*}

\cfang{Realizing such a pipelining schedule depends on how well the CGRA mapper can see and exploit these \jyincamrdy{opportunities for overlap}.} 
Existing recurrence-aware mappers \cite{park2008ems,oh2009rams,juneja2026compose} \jyincamrdy{already} attempt to minimize the \jyincamrdy{recurrence path} by prioritizing the scheduling of cyclic operations or by physically chaining PEs together. 
\cfang{However, such approaches generally schedule a dataflow graph (DFG) representation fixed in advance, without jointly exposing \mv{\jyincamrdy{other}} multi-context CGRA operator timing structures to the mapping process.} 
\jyincamrdy{As a result, the recurrence bound of Fig.~\ref{fig:overview}(a) still limits the overall throughput. But as Fig.~\ref{fig:overview}(c) demonstrates}, a further pipelined multi-context CGRA allows successive loop iterations to overlap across different configuration contexts, \jyincamrdy{hence launching workloads more frequently than the recurrence interval bound}. \jyincamrdy{
Yet, the resulting schedule still gives up 57\% of the ideal throughput than the ideal pipelining in Fig.~\ref{fig:overview}(b) that fully exploits interleaving and operator pipelining over both the multi-cycle operators and recurrence feedback paths.
Therefore, efficient mapping must jointly account for the recurrence path, the operator latency, the inter-tile routing delay, and other resource bounds are all accounted for.} 

\jyin{These observations motivate a broader shift in CGRA mapping: the dominant temporal bottlenecks and hardware timing structures should both shape the DFG representation and the schedule explored by the mapper. Hence, we present \system, a recurrence-aware temporal mapper for multi-cycle multi-context CGRAs, with the following contributions:}
\begin{enumerate}
    \item \textbf{Recurrence-aware loop-carried dependence analysis \cfang{(Sec.~\ref{sec:methodology_dfg_formation})}.} 
    Using hardware latency and loop-structure information, CONFERM identifies dataflow bottlenecks that limit either steady-state throughput or 
    \cfang{multi-cycle compute paths,}
    including recurrences and long-latency feed-forward paths. 
    Guided by these bottlenecks, it 
    identifies subgraphs to prioritize during scheduling,
    rather than relying on scheduling over a \cfang{DFG representation fixed in advance.} 
    \item \textbf{Multi-cycle-aware pipeline modeling and modulo forecast \cfang{(Sec.~\ref{sec:methodology_pipeline_forecast})}.} CONFERM unifies two temporal scales: 1) the operator-level pipeline schedule, defined by latency, launch interval, and pipeline occupancy, and 2) the CGRA-level modulo schedule, defined by initiation interval and inter-iteration overlap. Using a unified cycle-level architecture model, it forecasts the achievable initiation interval, temporal interleaving, and buffering demand for each loop-carried bottleneck before physical mapping. 

    \item \textbf{Bottleneck-first mapping with control-word folding \cfang{(Sec.~\ref{sec:methodology_mapping})}.} 
    CONFERM gives \mv{priority to} the forecasted bottlenecks during modulo scheduling on the target CGRA, while striving to maintain regular cycle offsets for hardware resource reservations across temporally interleaved iterations. 
    This regularity \jyincamrdy{enables control-word folding} that lets the emitted CGRA control sequence repeat at a shorter cadence than the mapping-stage initiation interval.
     
\end{enumerate}

\providecolor{revieworange}{RGB}{230,110,0}

\section{Background and Motivation}
\label{sec:motivation}

\subsection{Temporal Bottlenecks in Modulo Scheduling} 
\label{sec:motivation_bottlenecks}
\label{sec:bg_recurrence}
\jyin{A CGRA maps a loop DFG onto compute, routing, and storage resources while overlapping iterations. Under modulo scheduling~\cite{rau1994ims}, \cfang{operations from successive iterations are interleaved in time so that different iterations can be in flight together, with iterations starting every II cycles. This is illustrated by the four-cycle spacing between launches in Fig.~\ref{fig:overview}(a)}.}
\cfang{A smaller II lets more iterations overlap, and hence CGRA mappers seek to push II as low as the hardware allows.}

\jyincamrdy{Loop-carried dependence cycles impose a lower bound called the recurrence-constrained minimum initiation interval (RecMII)~\cite{mei2003modulo,park2006mge}.}
\cfang{A loop-carried dependence carries a value into a later iteration, and when it closes into a cycle with other DFG edges, the nodes on that cycle form a recurrent strongly connected component (SCC).}
\cfang{
\jyincamrdy{Over all recurrence cycles, RecMII is the maximum of the total cycle latency divided by the total iteration distance, rounded up to the nearest integer.}
The feedback path in Fig.~\ref{fig:overview}(a) is one such cycle, giving the four-cycle RecMII marked there.}

\cfang{The minimum II is also bounded by resource demand, denoted as the \textit{\textbf{res}}ource-constrained minimum II (ResMII). ResMII takes the largest ratio, over all resource types, of how many times that resource is needed per iteration to how many instances the array provides. In Fig.~\ref{fig:overview}(c), the kernel's seven operations on four PEs give a ResMII of two cycles.}
\cfang{The temporal bottleneck is therefore bounded as
\begin{equation}
    \text{minimum achievable II} \geq \max(\text{RecMII}, \text{ResMII}).
    \label{eq:state_update}
\end{equation}
}

\vspace{-1.5em}

\begin{figure}[t]
  \centering
  \includegraphics[width=\columnwidth]{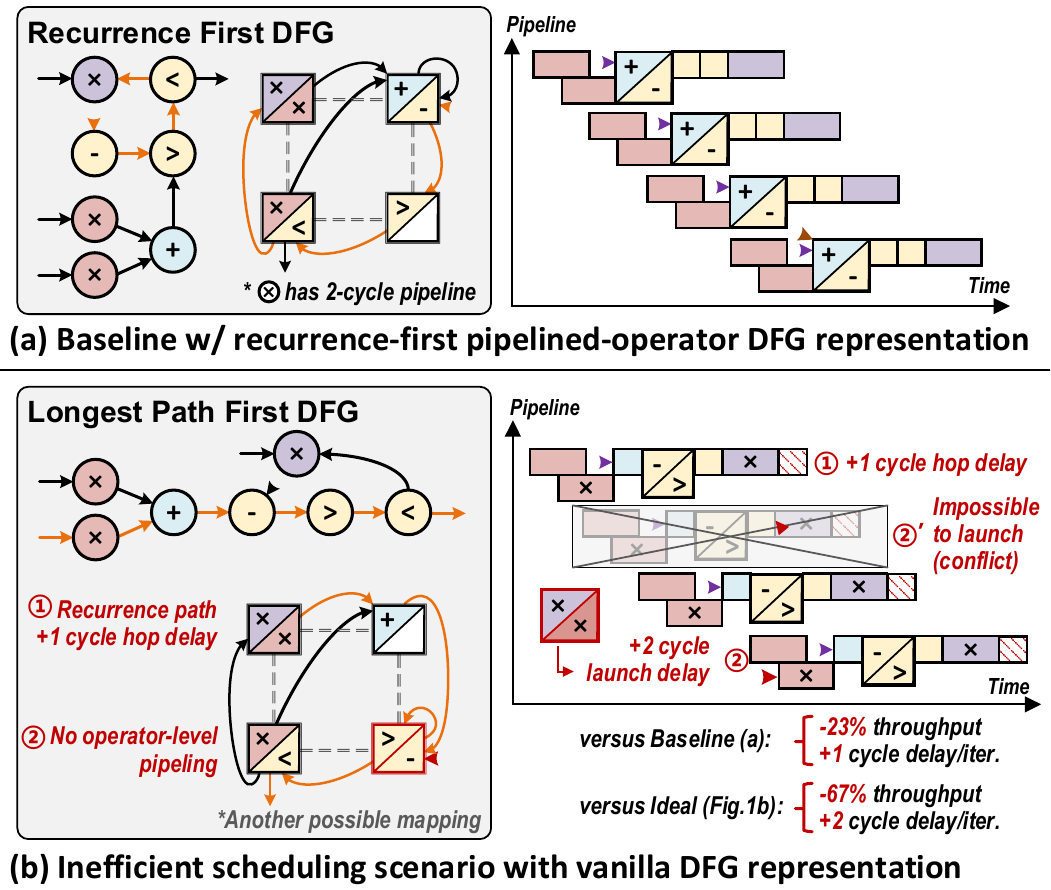}
  \caption{Effects of the mapper-visible DFG representation on mapping quality for the IIR kernel~\cite{oppenheim1989discrete}. Compared to the \textbf{(a)} baseline DFG annotated with the recurrence-path priority and operator-pipeline capability, the \textbf{(b)} naive DFG representation along the longest delay path can result in inefficient scheduling because of \circled{1} longer recurrence path delay and \circled{2} lack of pipeline information available to the mapper. The baseline case is the same as Fig.~\ref{fig:overview}(c). }
  \label{fig:fused-overhead}
\end{figure}

\subsection{Multi-cycle Multi-Context CGRA} \label{sec:motivation_cgra}
\label{sec:bg_architecture}
\cfang{The RecMII and ResMII bounds depend on what the target hardware actually provides.}
\jyin{Fig.~\ref{fig:arch}(a) shows a \(4\times4\) multi-context CGRA whose cores contain pipelined single-instruction, multiple-data (SIMD) compute units, local and mesh first-in, first-out (FIFO) storage, crossbars, and private instruction and constant buffers~\cite{yin2026stella}. 
\cfang{Each core, shown in Fig.~\ref{fig:arch}(b)}, repeatedly executes instruction contexts from a multi-entry buffer, and thus operations, routes, and buffered values occupy hardware execution slots at specific modulo phases. 
\cfang{Fig.~\ref{fig:arch}(c) shows three structures, elastic FIFOs, multi-cycle compute units, and circular read-only buffers, that let a value stay in the hardware for more than one cycle before it has to be consumed.}}

\jyin{\cfang{Among these, a multi-cycle operator is defined by its result latency and minimum launch interval. The launch interval can be shorter than the result latency \mv{by means of} hardware \mv{pipelining}.}
\cfang{In Fig.~\ref{fig:overview}(a), the multiplier has a result latency of two cycles, but if its launch interval is one cycle, it can start a new multiplication before the previous one finishes. As a result, independent recurrent streams and independent iterations of an acyclic path can occupy different temporal slots of the same hardware unit concurrently.}
Yet, the mapper still has to ensure feasibility under the place-and-route resources and control-memory limits.}

\subsection{Mapper-Visible Representation} \label{sec:motivation_dfg_representation}

\cfang{Whether a CGRA mapper can exploit the pipelining resources depends on the DFG representation it is given. Fig.~\ref{fig:fused-overhead}(a) and Fig.~\ref{fig:fused-overhead}(b) map the same IIR kernel under two different representations. In Fig.~\ref{fig:fused-overhead}(a), the multiplier's two-cycle pipeline is exposed to the mapper as a single operator, and the resulting schedule stays close to the RecMII and ResMII bounds of Fig.~\ref{fig:overview}. In Fig.~\ref{fig:fused-overhead}(b), the same computation is instead represented along its longest path, hiding the multiplier's internal pipelining from the mapper. Without this visibility, the mapper cannot issue a second multiplication before the first one completes, forcing an impossible pipeline stage and a two-cycle launch delay. The recurrence path itself also \jyincamrdy{incurs an additional cycle of routing delay} from routing the value between separate tiles, further lengthening the path that sets RecMII. Fig.~\ref{fig:fused-overhead} reports the resulting cost: 23\% less throughput and one extra cycle of delay per iteration compared to the baseline representation of Fig.~\ref{fig:fused-overhead}(a), and 67\% less throughput and two extra cycles of delay per iteration compared to the ideal pipeline of Fig.~\ref{fig:overview}(b).}

\jyin{Existing recurrence-aware CGRA mappers~\cite{park2008ems,oh2009rams,juneja2026compose} prioritize cyclic operations but generally schedule the DFG they receive.
Other works~\cite{tan2023vecpac,tan2020opencgra,qin2025picachu,prabhakar2017plasticine,park2009cgraexpress} \cfang{perform fusion, vectorization, or pattern-level execution on the DFG, but treat this representation choice separately from the loop-carried timing that recurrence-aware scheduling addresses.}
\cfang{These observations \jyincamrdy{motivate a mapper that addresses both bottlenecks} by forming DFG representations that expose what the target CGRA's operators can pipeline.}
}

\section{Methodology of \system} \label{sec:methodology}

\begin{figure*}[!t]
  \centering
  \includegraphics[width=\linewidth]{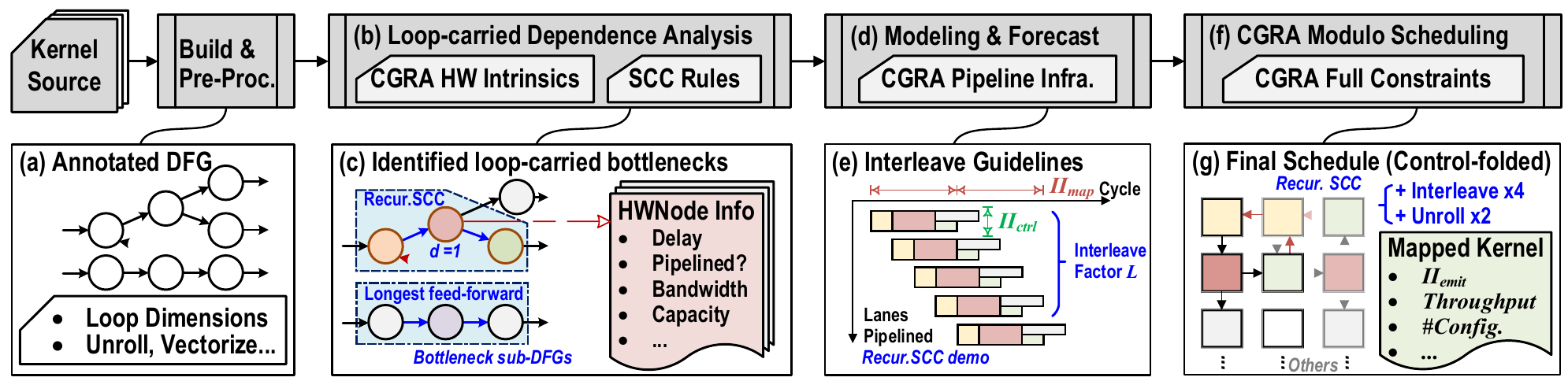}
  \caption{The \system CGRA mapping flow. Starting from the LLVM-based build of kernel dataflow graph with loop structure information \textbf{(a)}, the CGRA hardware operator intrinsics are annotated into the DFG and trigger rule-based strongly connected component detection \textbf{(b)}. The resulting DFG \textbf{(c)} carries identified loop-carried bottlenecks including the recurrent SCC and longest feed-forward chain. Then the multi-cycle pipeline modeling and modulo forecasting stage \textbf{(d)} generates mapping guidelines \textbf{(e)} targeting a low initiation interval (II) via interleaving and pipelining the bottlenecks under CGRA pipeline infrastructure constraints. Finally, a heuristic-based CGRA mapper backend \cite{tan2020opencgra} conducts the actual modulo scheduling and prioritizes forecasted guidelines \textbf{(f)}. The mapped schedule \textbf{(g)} features higher throughput, lower control overhead, and a shorter emitted control period. }
  \label{fig:flow}
\end{figure*}

\subsection{Flow Overview and Preparation} \label{sec:methodology_flow_overview}

As shown in Fig.~\ref{fig:flow}, \system is built on top of the heuristic OpenCGRA mapper~\cite{tan2020opencgra} \jyincamrdy{which extends} classical modulo scheduling~\cite{rau1994ims} procedure. That is, \jyincamrdy{kernel operations reserve CGRA compute and routing resources within an II-cycle modulo window}, including both the computation and data-routing requests.
On top of this, \system further implements a rewrite-forecast-confirm flow to leverage the pipeline opportunities in the target multi-cycle multi-context CGRA in Fig.~\ref{fig:arch} to boost the throughput via loop-carried schedules.

Before entering the main processing steps, \system builds the standard kernel DFG via frontend compilation and preprocessing.
The target application workload is written in C and compiled into LLVM intermediate representation (IR). 
Then, pattern-based graph-rewrite passes \jyincamrdy{transform the initial} LLVM IR DFG to align with the target hardware. To match the target CGRA's supported operators in Fig.~\ref{fig:arch}(b), unsupported arithmetic operators are lowered to CGRA intrinsics (e.g. funnel-shift as shift-shift-or, compare-select as min-max) while common computation patterns are fused into advanced operators (e.g. multiply-accumulate-quantization as \textit{cgraFMA}).
Meanwhile, hardware primitives are annotated into each DFG node especially for the delay and pipeline capability (Fig.~\ref{fig:flow}(a), Fig.~\ref{fig:iir_transform}(a)). 
For fairness, this preparation step, \mv{together with the CGRA hardware architecture parameters, is} shared across all evaluated mapping methods in this paper.

\subsection{Recurrence-aware Loop-carried Dependence Analysis}  \label{sec:methodology_dfg_formation}

\mv{Starting from} the rewritten DFG with hardware latencies and loop dimensions annotated, \system carries out the loop-carried dependence analysis to \jyincamrdy{identify bottleneck subgraphs and determine whether they contain recurrences.}
This is achieved by extracting the iteration-distance labels derived from LLVM IR loop information. From the extraction, every edge in the DFG gets a distance $d$. A loop-carried data edge entering the loop header is assigned distance $d=1$. The control edge along the back edge is also $d=1$. Everything else gets $d=0$. 

\jyincamrdy{Before recurrence analysis}, \system offloads induction-variable updates and loop control to the peripheral memory load/store units in Fig.~\ref{fig:arch}(a).
Specifically, the memory addressing arithmetic and loop-control recurrence (add-cmp-br-phi) are detected and folded into an affine address pattern, \mv{which is} then completely offloaded into the memory IO units as \textit{\textbf{streamLoad}}/\textit{\textbf{streamStore}} operators. 
These offloaded recurrences are excluded from bottleneck analysis.
After this \textit{\textbf{stream}} conversion, some phi nodes may remain which denote delay registers that carry data between iterations. Hence, a further sweep absorbs these phi nodes by incorporating their carry depths as the extra distance $\Delta d$ into the adjacent edge. 

\begin{figure}[!t]
  \centering
  \includegraphics[width=\columnwidth]{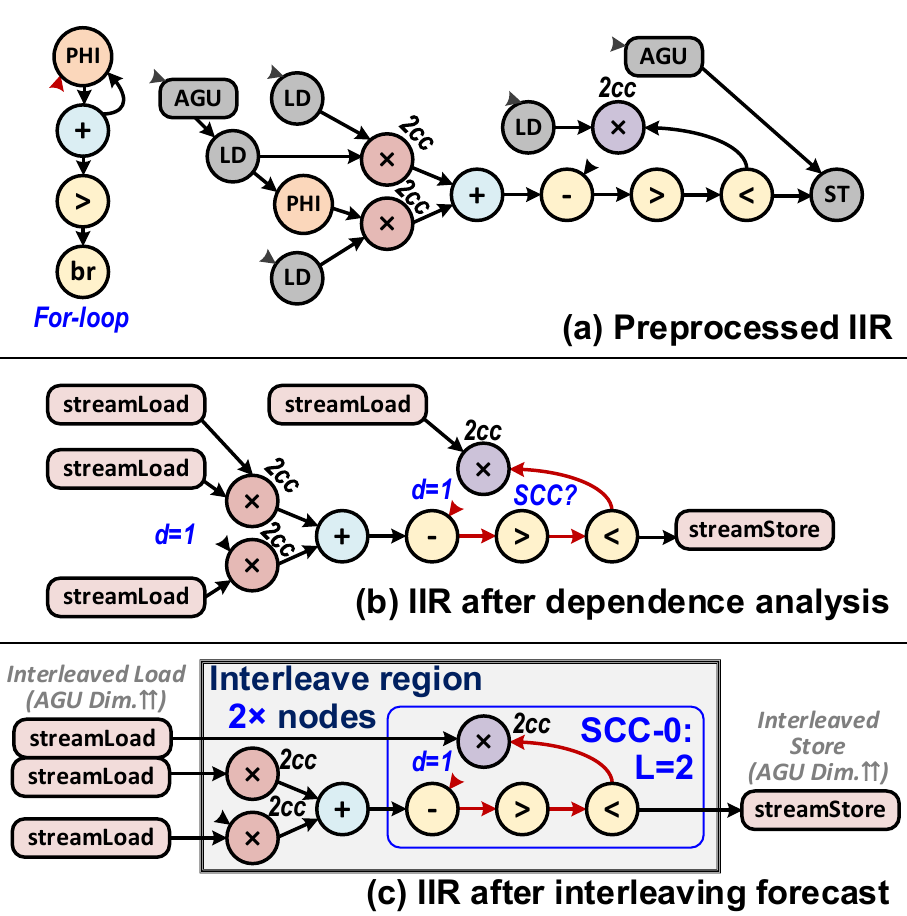}
  \caption{DFG transformations across \system stages using the IIR kernel~\cite{oppenheim1989discrete}: \textbf{(a)} the preprocessed DFG with operators lowered and fused towards the target CGRA intrinsics; \textbf{(b)} the bottleneck-annotated DFG after the stream memory IO conversion, edge distance merging, and loop-carried distance analysis; \textbf{(c)} the DFG with interleaving factor guidelines after the step of multi-cycle pipeline modeling and modulo forecast.  }
  \label{fig:iir_transform}
\end{figure}

After this extra pre-processing, the bottleneck structure is located by detecting $d\geq1$. 
Once identified, we use Tarjan’s algorithm~\cite{tarjan1972depth} to recursively check strongly connected components (SCCs) for recurrence (Fig.~\ref{fig:flow}(b)).
\jyincamrdy{If no recurrence is identified}, \system falls back to using the longest-delay chain of the DFG as the bottleneck.
Together, the outcomes of these categories form \jyincamrdy{the subgraphs targeted for optimization} (Fig.~\ref{fig:flow}(c) and~\ref{fig:iir_transform}(b)) for further steps.

\subsection{Multi-cycle-aware Pipeline Modeling and Modulo Forecast}  \label{sec:methodology_pipeline_forecast}

Using the annotated DFGs and identified loop-carried dependence bottlenecks, \system brings in further hardware pipelining infrastructure parameters to perform the multi-cycle-aware pipeline modeling and modulo forecast (Fig.~\ref{fig:flow}(d)).

First, CONFERM recalculates the modulo-scheduling bounds. Let $N$ be the number of compute nodes after preprocessing (excluding \textit{\textbf{stream}}/\textit{\textbf{phi}} nodes), $F$ be the number of CGRA PEs, and let a bottleneck chain $C$ have total latency $\Delta_C = \sum_{v\in C}\ell(v)$ over its nodes $v$ and total distance $d_C = \sum_{e\in C} d(e)$ over its edges $e$.
The updated classical modulo-scheduling bounds are
\begin{equation}
\mathrm{ResMII} = \left\lceil \frac{N}{F} \right\rceil, \qquad
\mathrm{RecMII} = \max_{C} \left\lceil \frac{\Delta_C}{d_C} \right\rceil .
\label{eq:resmii}
\end{equation}

Then, per sub-CU in category $c$ (ALU, MAC, LUT in Fig.~\ref{fig:arch}(b)), computation nodes are pooled to accumulate the issue cycle occupancy 
and set sub-CU resource boundaries beyond the aggregate ResMII.
Across the target DFG with $n_c$ nodes of category $c$ occupying $i_c$ issue cycles each on $F$ PEs, the sub-CU bound and the per-iteration control period are
\begin{equation}
b_c = \left\lceil \frac{n_c\, i_c}{F} \right\rceil, \qquad
\mathrm{II}_{\mathrm{ctrl}} = \max\bigl(1,\ \mathrm{ResMII},\ \max_c b_c\bigr) ,
\label{eq:iiinstr}
\end{equation}
with $\mathrm{II}_{\mathrm{ctrl}}$ denoting the minimal iteration period on the target hardware, in other words the maximal achievable throughput scenario.
Using this bound, CONFERM interleaves recurrent and feed-forward DFG regions into loop-carried pipelines that improve aggregate throughput by interleaving independent streams as showcased in Fig.~\ref{fig:overview}(b-c) and Fig.~\ref{fig:fused-overhead}.
That is, a loop-carried body whose latency $\Delta_C>d_C \cdot\mathrm{II}_{\mathrm{ctrl}}$ cannot run one iteration per $\mathrm{II}_{\mathrm{ctrl}}$ period. 
\jyincamrdy{Accordingly}, the modulo forecast unit (Fig.~\ref{fig:flow}(d)) applies C-slow interleaving~\cite{leiserson1991retiming} estimation for the body with $L$ independent iterations, where
\begin{equation}
L =
\max_{C \in \mathcal{B}}
\left\lceil
\frac{\Delta_C}
{\max(1,d_C)\cdot\mathrm{II}_{\mathrm{ctrl}}}
\right\rceil.
\label{eq:lint}
\end{equation}
\jyincamrdy{
Here, \(\mathcal{B}\) contains the detected recurrence cycles or, if none exist, the longest feed-forward path where \(d_C=0\). 
The loop-carried launch spacing in the latter case is constrained by operator launch intervals and resource capacities, while path latency affects in-flight hardware occupancy without imposing a feedback constraint.
}
The $L$-interleaving resources come from independent computation in different loop phases (e.g. loop unrolling) or batched iterations (Fig.~\ref{fig:overview}(b)).
Hence, as Fig.~\ref{fig:flow}(e) and Fig.~\ref{fig:iir_transform}(c) illustrates, the next-step mapping II starts from 
\begin{equation}
\qquad \mathrm{II_{map}} = L \cdot \mathrm{II}_{\mathrm{ctrl}} .
\label{eq:ii_map}
\end{equation}
Therefore, the lifetime of each operation and value inside this interleaved DFG bundle is prolonged beyond $II_{ctrl}$, which means hardware resources are required to support an estimated number of simultaneously live values given by
\begin{equation}
B_e = \left\lceil \frac{\Delta_e}{\mathrm{II}_{\mathrm{ctrl}}} \right\rceil
\quad\text{times}.
\label{eq:be}
\end{equation}
\jyincamrdy{The mapper checks this estimate against} the CGRA FIFO depth and CU pipeline capacity during the mapping by the mapper's admission rules \cite{tan2020opencgra}. If constrained by CGRA resources, the final mapped interleaving density would be lower than the forecast $L$ as illustrated by examples in Fig.~\ref{fig:overview}(c) and Fig.~\ref{fig:fused-overhead}(b).

\subsection{Bottleneck-first Mapping with Control-word Folding}  \label{sec:methodology_mapping}

\begin{figure}[!t]
  \centering
  \includegraphics[width=\columnwidth]{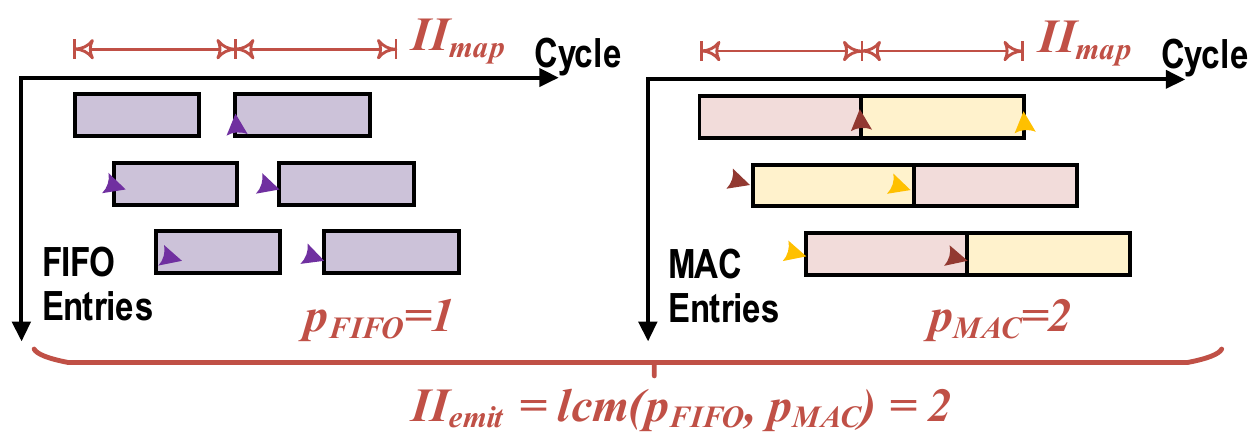}
  \caption{Control-word folding in a FIFO–MAC example with $II_{map}=4$. The recorded resource occupancy and mapped activities show a minimal repetition period $p_i$ of 1 for the FIFO and 2 for the MAC unit. Hence the final $II_{emit}=2$, which is 50\% shorter than the mapping period $II_{map}=4$.}
  \label{fig:ctrl_folding}
\end{figure}

In this stage (Fig.~\ref{fig:flow}(f)), \system carries out the actual placement and routing of the target DFG with the prioritized loop-carried structure and forecasted interleaving schemes.

Because of the incompleteness and sequential mapping behavior of the heuristic mapper backend \cite{tan2020opencgra}, \system generates multiple candidates that explore different mapping-priority orders (e.g. the recurrence-first and longest-first choices in Fig.~\ref{fig:fused-overhead}).
Further, if the mapping attempt fails at a certain $II_{map}$, the mapper increases $II_{map}$ until it reaches the control memory capacity constraint.
In this case, there is a feedback check for the new $II_{map}$ target by the loop-carried forecast unit (Fig.~\ref{fig:flow}(d)). 
If the interleaving unit has \jyincamrdy{atomic interleaving granularity}, which is common in time-varying recurrence, 
candidate $II_{map}$ periods that violate this granularity are skipped
Finally, if the feasibility check passes, the updated interleaving factor $L$ and resource pressure ratio $B_e$ are updated via Eqs.~\ref{eq:ii_map} and \ref{eq:be}, respectively.

Once a candidate mapping is validated, the actual configuration overhead would be derived from the $II_{map}$ via the control-word folding search.
Because interleaving produces regular control patterns, the unique control sequence needed to represent one mapping period can be much smaller than $II_{map}$\mv{, hence saving} configuration overhead without specific control compression techniques.
For example, a systolic schedule can repeat every cycle even when a computation spans many cycles~\cite{moldovan1983design}.
For multi-context CGRAs with VLIW-like control (very long instruction words, \cite{rau1994ims}), the modulo folding needs to be carried out for each hardware instance first, mainly including the compute units, crossbar and FIFO array read/write indexing. For hardware instance $i$ with its scheduled control word sequence $w_i(t)$, the smallest period reproducing it is calculated by
\begin{equation}
\begin{aligned}
p_i
&= \min\{\, p : p \mid \mathrm{II_{map}},\ w_i(t) = w_i(t+p)\ \forall t \,\}.
\end{aligned}
\label{eq:collapse}
\end{equation}
At the same time, the actual interleaving factor $L_{emit}$ is confirmed by the mapper. Together with the source code unrolling factor $u$, we can derive respectively the delivered II period (Fig.~\ref{fig:ctrl_folding}), throughput and configuration overhead:
\begin{equation}
\begin{aligned}
\mathrm{II}_{emit}
&= \operatorname{lcm}_i\, p_i ,\\
\mathrm{Throughput} &= \frac{u\cdot L_{emit}}{II_{map}},\\
N_{\mathrm{configuration}} &= \sum_i p_i.
\end{aligned}
\label{eq:metrics}
\end{equation}
We report $\mathrm{Throughput}$ as source loop iterations completed per cycle, so that every mapping technique, whether a plain modulo schedule, an unrolled body, or a temporally interleaved lane, is measured by the same unit of work as written in the kernel source, independent of how many operations, control words, or PEs each technique spends to deliver it.
Finally, \system selects the candidate with the highest throughput, breaking ties by the smallest configuration footprint.

\section{Evaluation Methodology}

\subsection{Evaluation Setup}

The \system features (Fig. \ref{fig:flow}) are implemented as custom LLVM passes on top of the baseline OpenCGRA mapper \cite{tan2020opencgra}. The multi-context CGRA hardware modeling (Fig. \ref{fig:arch}) inside the mapper is derived from a silicon-proven vectorized CGRA SoC STELLA \cite{yin2026stella}, especially the pipeline resource infrastructure and the cycle-level compute unit delays. The CGRA mesh is parameterizable with 4$\times$4 as the default size.
The operator vectorization is normalized in this work by applying the same SIMD hardware operator support across all mapper solutions.

To evaluate the loop-carried mapping performance, we compare CONFERM against COMPOSE \cite{juneja2026compose} and PICACHU \cite{qin2025picachu}. 
To ensure a fair comparison under a unified infrastructure, we reproduce both approaches within the CONFERM flow and the same hardware model by disabling passes unsupported by the corresponding prior work, and refer to these configurations as COMPOSE-like and PICACHU-like, respectively. For the ablation study, we additionally evaluate the original OpenCGRA \cite{tan2020opencgra} as Vanilla CGRA, and CONFERM-Recur.Only, a reduced CONFERM configuration with support for non-recurrence loop-carried dependencies disabled.
Table~\ref{tab:eval_targets} summarizes the features supported by each configuration.

\begin{table}[t]
    \centering
    \caption{Feature coverage of the evaluated mapping solutions}
    \label{tab:eval_targets}
    \setlength{\tabcolsep}{1.6pt}
    \renewcommand{\arraystretch}{1.04}
    \begin{tabular}{lccc}
        \toprule
        \textbf{Configuration}
        & \shortstack{\textbf{Multi-cycle}\\\textbf{Pipelining}}
        & \shortstack{\textbf{Loop-carried}\\\textbf{Recurrence}}
        & \shortstack{\textbf{Loop-carried}\\\textbf{Feed-forward}} \\
        \midrule
        Vanilla CGRA~\cite{tan2020opencgra}          & --         & --         & --         \\
        PICACHU-like~\cite{qin2025picachu}         & \checkmark & --          & --         \\
        COMPOSE-like~\cite{juneja2026compose}         & --    & \checkmark         & \checkmark         \\
        \hline
        CONFERM-\textit{Recur.Only}    & \checkmark & \checkmark & --         \\
        CONFERM               & \checkmark & \checkmark & \checkmark \\
        \bottomrule
    \end{tabular}
\end{table}

\begin{table}[t]
  \centering
  \caption{Target benchmark set and longest-chain characterization}
  \label{tab:benchmarks}
  \setlength{\tabcolsep}{1.6pt}
  \renewcommand{\arraystretch}{1.04}
  \resizebox{\columnwidth}{!}{\begin{tabularx}{\columnwidth}{c l L{7em} c C{2.2em} C{2.2em} C{6em}@{}}
    \toprule
    \multirow{2}{*}{\textbf{Category}} &
    \multirow{2}{*}{\textbf{Kernel}} &
    \multirow{2}{*}{\textbf{Description}} &
    \multirow{2}{*}{\textbf{\#Nodes}} &
    \multicolumn{3}{c}{\textbf{Longest Path}} \\
    \cmidrule(lr){5-7}
    & & & & Fwd. & Rec. & \#MC-Nodes \\
    \midrule
    \multirow{4}{*}{\shortstack{\textbf{MC}\\\textbf{Recur-}\\\textbf{rent}}}
  & IIR    & IIR biquad filter       & 18 &  9 &  4 & 2 \\
      & LogSumExp  &  Online Norm.      &  24 & 12 & 2  & 2 \\
      & FlashAttn.    & Online softmax      & 56 & 16  & 2 & 3 \\
      & MambaSSM  &  Selective Scan        & 51 & 14 & 2 & 4 \\
    \midrule
    \multirow{3}{*}{\shortstack{\textbf{LL}\\\textbf{Bitwise}}}
      & Viterbi & Viterbi decoding        & 23 &  7 & 8 & - \\
      & Dither  & Image dithering         & 12 &  6 &  3 & - \\
      & TinyDES & DES encryption          & 14 &  7 &  - & - \\
    \midrule
    \multirow{3}{*}{\shortstack{\textbf{MC}\\\textbf{Feed-}\\\textbf{forward}}}
      & FFT    & Fourier transform        & 29 &  7 &  - & 2 \\
      & GeLU    & Gaussian Act.    & 17 & 13 &  - & 6 \\
      & SwiGLU  & Swish gating        & 16 &  9 &  - & 3 \\
    \bottomrule
  \end{tabularx}}

  \vspace{2pt}
  \parbox{\columnwidth}{\fontsize{7.5}{8.0}\selectfont
  \textbf{\dag} MC = multi-cycle; LL = low-latency, i.e. multi-cycle-poor; Fwd. = Feed-forward; Rec. = Recurrence; Act. = Activation. Kernels are profiled with unroll-1.}
\end{table}

\subsection{Benchmark Kernels}

Table~\ref{tab:benchmarks} summarizes the benchmark suite, which comprises 10 kernels across three categories designed to exercise the mapper capabilities identified in Table~\ref{tab:eval_targets}: 1) \textbf{Multi-Cycle Loop-carried}: four kernels containing loop-carried recurrences whose dependence paths include multi-cycle operators~\cite{dao2022flashattention,gu2024mamba}; 2) \textbf{Low-Latency Bitwise}: three benchmarked kernels adopted from COMPOSE~\cite{juneja2026compose}, containing only single-cycle operators; 3) \textbf{Multi-Cycle Feed-forward}: three \jyincamrdy{feed-forward} kernels adopted from PICACHU~\cite{qin2025picachu}, whose feed-forward DFG paths contain heavier multi-cycle computation than their recurrence paths.

Table~\ref{tab:benchmarks} also reports the node-level characteristics of the raw kernel DFGs. During the evaluation, the effective DFG complexity may decrease after preprocessing to match the CGRA hardware intrinsics (Fig.~\ref{fig:arch}). 
Importantly, the same preprocessed kernel DFGs, \mv{as well as identical hardware capabilities,} are used as the common input to all evaluated mapper \mv{variants} for fairness.

\section{Experimental Results}

\begin{figure*}[!htbp]
  \centering
  \includegraphics[width=\linewidth]{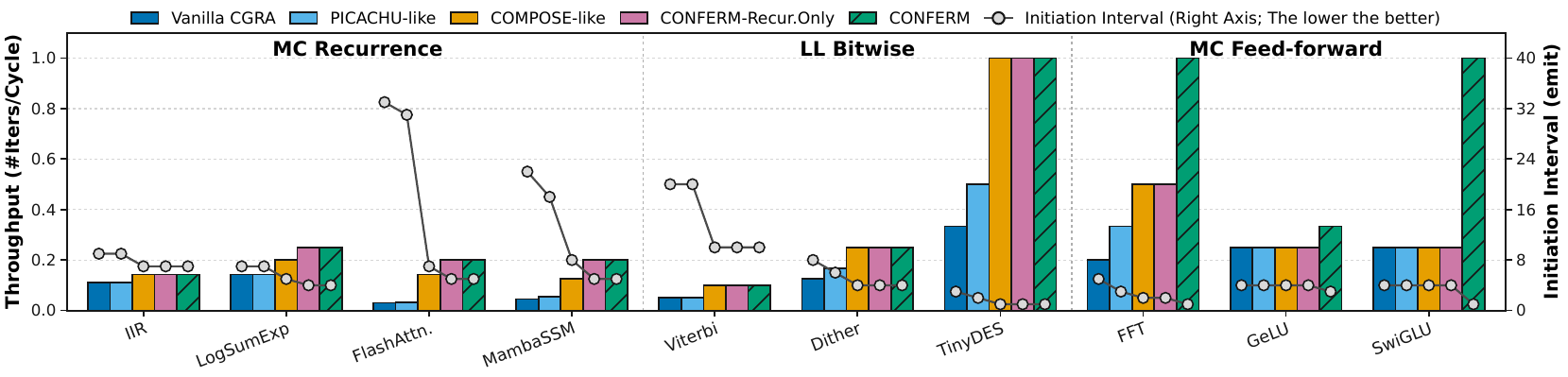}
  \caption{The end-to-end mapping quality comparison of CONFERM against other mapping solutions in Table.~\ref{tab:eval_targets}, shown by the achieved throughput (left axis) and the $II_{emit}$ initiation interval (right axis). The kernels used are defined in Table.~\ref{tab:benchmarks} with an unrolling factor of 1. \jyincamrdy{The target is a default \textbf{4$\times$4} CGRA.}}
  \label{fig:res_e2e}
\end{figure*}

\begin{figure*}[!htbp]
  \centering
  \includegraphics[width=\linewidth]{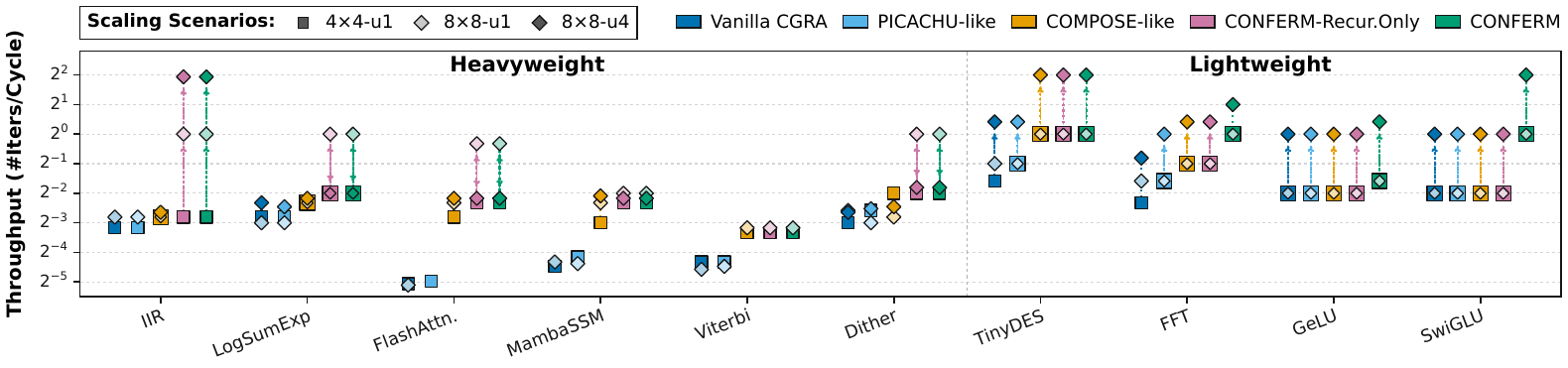}
  \caption{\jyincamrdy{The scaling-up evaluation that evaluates mapping quality of CONFERM under hardware scaling and loop unrolling. The CGRA hardware is scaled up to \textbf{8$\times$8 }and tested with unroll-\textbf{1/4} kernels. The throughput trajectories show how each mapping technique behaves differently as array size and unrolling factor change. Kernels are re-categorized into \textit{\textbf{Heavyweight}} and \textit{\textbf{Lightweight}} groups by the mapping complexity. A missing marker indicates a failed mapping attempt for the corresponding kernel and configuration.}}
  \label{fig:res_e2e_scale_up}
\end{figure*}

\subsection{End-to-end Mapping Quality}

Figure~\ref{fig:res_e2e} presents the end-to-end mapping results of the mapping solutions in Table.~\ref{tab:eval_targets} under the same software and hardware configuration (4$\times$4 CGRA tiles and unroll-1 kernels).

Across all ten benchmarked kernels, CONFERM proves its efficacy by jointly considering recurrences and multi-cycle operator pipelines during the modulo mapping. It achieves geometric-mean throughput speedups of \textbf{2.71$\times$}, \textbf{2.33$\times$}, \textbf{1.50$\times$} over Vanilla CGRA~\cite{tan2020opencgra}, PICACHU-like~\cite{qin2025picachu}, and COMPOSE-like~\cite{juneja2026compose} as state-of-the-art works, together with an average reduction in the emitted control period of 57.4\%, 52.1\%, and 28.6\%, respectively. 
CONFERM's strengths are most apparent by the \textbf{MC Recurrence} group which contains the most complex kernels that have both the recurrence and multi-cycle operators.

For the \textbf{\textit{LL Bitwise}} group (Viterbi, Dither, TinyDES), the proposed CONFERM and CONFERM-Recur.Only achieve the same mapping outcomes as COMPOSE-like. This is expected because these kernels contain only single-cycle operators where recurrence-aware prioritization accounts for the gains in the recurrent kernels, effectively eliminating the optimization opportunity associated with multi-cycle operator pipelining. 
Compared with the recurrence-unaware methods (Vanilla and PICACHU-like), the throughput boosts are \textbf{2.29$\times$} and \textbf{1.82$\times$}, respectively. 
These results show the value of considering recurrence structure when assigning mapping priorities for heuristic mapping.

For the \textbf{\textit{MC Feed-forward}} group (FFT, GeLU, SwiGLU), the mapping that targets regular inter-iteration offsets with control-word folding (Sec.~\ref{sec:methodology_mapping}) improves throughput for CONFERM versus mappers without these scheduling constraints. The multi-cycle operator pipelining of PICACHU-like mapper is not fully exploited here because the kernel unrolling factor is 1. A follow-up experiment examines this effect in the next section.

\subsection{Kernel Parallelism Ablation}

Beyond mapping a single innermost loop body on the default CGRA, \jyincamrdy{hardware scaling and loop unrolling are the two} common practices to seek higher parallelism and throughput. Both increase the scheduling demands on the mapper, \jyincamrdy{especially for the recurrence scenarios}.

\begin{figure}[]
  \centering
  \includegraphics[width=\columnwidth]{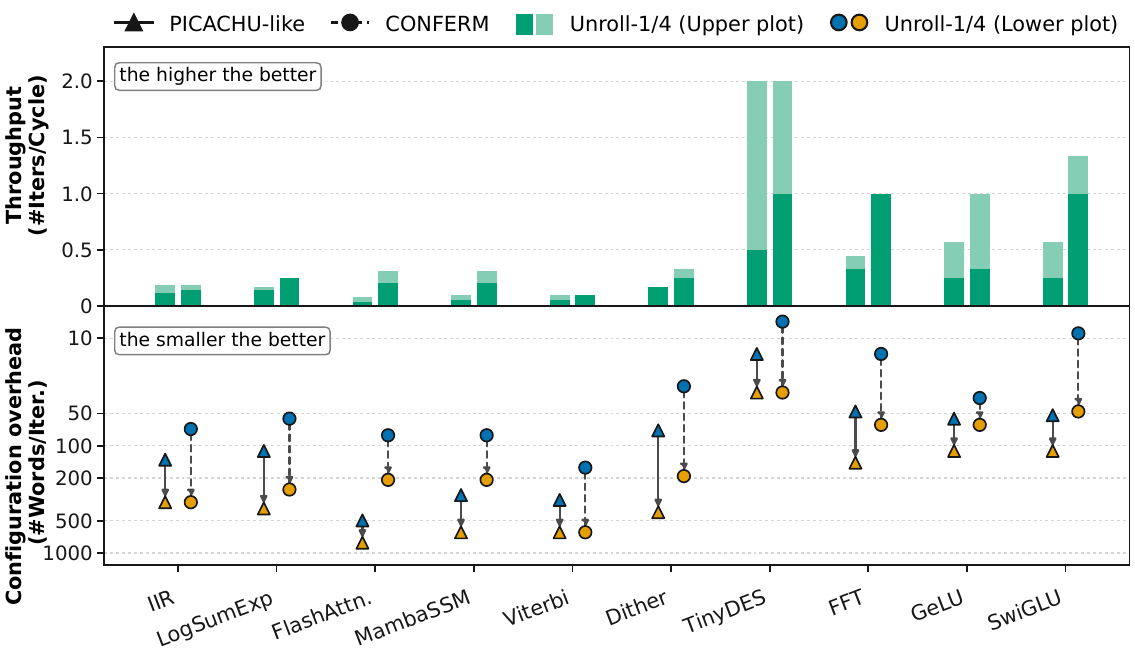}
  \caption{The comparison of mapping quality for unrolling factors of 1 and 4 between PICACHU-like and CONFERM. Together with the throughput metric, the configuration overhead by total control words launched per active CGRA core per iteration is evaluated to complement the emitted control-period metric. }
  \label{fig:unroll_vs_PICACHU}
\end{figure}

\jyincamrdy{
As Fig.~\ref{fig:res_e2e_scale_up} shows, scaling up the CGRA from 4$\times$4 to 8$\times$8 releases more scheduling resources for \textit{\textbf{Heavyweight}} kernels. Because the resource bound is lifted, most mappers achieve higher throughput, with particularly large gains for CONFERM because its loop-carried pipelining enforces more mapping constraints. 
Against Vanilla CGRA, PICACHU-like, and COMPOSE-like, CONFERM's geo-mean throughput speedups are further boosted to \textbf{7.04}×, \textbf{5.65}×, and \textbf{3.27}×, respectively.
Further, with loop unrolling enabled, \textit{\textbf{Lightweight}} kernels generally achieve higher throughput by leveraging the underutilized 8$\times$8 CGRA in unroll-1 scenarios. 
However, for \textit{\textbf{Heavyweight}} kernels with data-recurrence structures, unroll-4 creates $\sim$4$\times$ larger DFGs instead of four independent subgraphs.
Hence, their achieved mappings become resource-constrained or fail, where CONFERM is the most robust under these conditions. For IIR, a relatively simple recurrent kernel, CONFERM fully demonstrates its efficacy to boost achievable throughputs further by jointly leveraging the hardware scaling and loop unrolling.
}

To evaluate the detailed mapper behaviors further, we measure the configuration overhead accumulated from active CGRA cores per iteration, as an additional mapping quality metric to the flattened $II_{emit}$. 
\jyincamrdy{
We compare CONFERM against PICACHU-like, both supporting multi-cycle operator pipelining, on 4$\times$4 CGRA tiles with unroll-1/4 kernels.
}
Fig.~\ref{fig:unroll_vs_PICACHU} shows that over all ten kernels, CONFERM delivers a geometric-mean throughput boost of \textbf{2.3$\times$} over PICACHU-like with the unroll-1 setting. This advantage decreases to \textbf{1.8$\times$} with unroll-4 scenarios, where PICACHU-like catches up by mapping independent innermost loop bodies freely, \jyincamrdy{especially for \textit{\textbf{Lightweight}} kernels}.
Yet, the bottleneck-first mapping and control-word folding reduce control-word overhead by \textbf{64\%} and \textbf{44\%} respectively for unroll-1 and unroll-4 ablations, respectively. 
This benefits the energy efficiency on real hardware by saving the instruction fetching and decoding overhead.

\subsection{Mapper Convergence Speed versus the Mapping Quality}

\begin{figure}[]
  \centering
  \includegraphics[width=0.99\columnwidth]{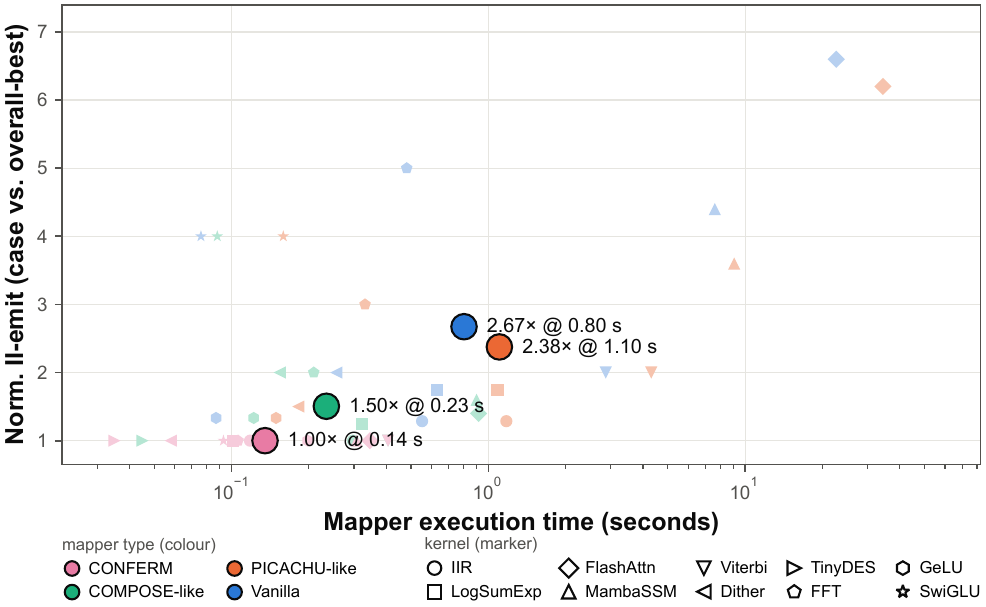}
  \caption{Mapper execution time in seconds (geo-mean over 10 runs per case) versus the achieved II-emit over all the benchmarked kernels and mappers. The scenario settings are the same as Fig.~\ref{fig:res_e2e}. The achieved II-emit is normalized to the best overall II-emit of the corresponding kernel. Then a final average execution time is computed for each mapper as the Pareto-frontier metric.}
  \label{fig:utilization}
\end{figure}

We also profile the execution time for each mapper to find a valid schedule for each kernel. 
For each case, we report the geometric mean of the mapping runtimes over ten runs.
As shown in Fig.~\ref{fig:utilization}, the recurrence-aware mappers (COMPOSE-like and CONFERM) generally converges faster in finding a valid mapping. 
This is reasonable because the detection of recurrence is applicable to 6 of 10 benchmarked kernels where the priority mapping pays off. 
Overall, CONFERM's joint consideration of loop-carried dependences and CU pipelines provides strong guidance that speeds up the CGRA mapping and achieves better overall mapping quality, thus pushing forward the Pareto frontier under the same heuristic mapping settings.
The mapping-runtime speedups over Vanilla CGRA, PICACHU-like, and COMPOSE-like are \textbf{5.71$\times$} (Vanilla), \textbf{7.85$\times$} (PICACHU-like), and \textbf{1.64$\times$} (COMPOSE-like), respectively.

\section{Conclusion}
This work presents CONFERM, a recurrence-aware temporal mapper for multi-cycle multi-context CGRA scheduling. 
For the identified bottlenecks in modern DSP and machine learning workloads, CONFERM carries out recurrence-aware loop-carried dependence analysis to prioritize recurrent subgraphs and long-latency chains of multi-cycle operators.
By combining operator-level pipeline schedule and CGRA-level modulo scheduling, CONFERM conducts bottleneck-first mapping with forecast-guided scheduling followed by control-word folding.
Across all 10 benchmarked kernels, CONFERM improves throughput by 2.18$\times$ over state-of-the-art mappers and its regular iteration offsets shorten the emitted initiation interval by 46\%. 
Meanwhile, CONFERM's mapper pass also converges 5.07$\times$ faster on the same heuristic mapper backend.

\end{document}